\documentclass[preprint,showpacs,preprintnumbers,amsmath]{revtex4}
\usepackage{graphicx}
\usepackage{dcolumn}
\usepackage{bm}

\begin{document}

\title{Entanglement dynamics of two independent Jaynes-Cummings atoms without
rotating-wave approximation}
\author{Qing-Hu Chen$^{1,2}$, Tao Liu$^{3}$, Yuan Yang$^{1}$, and Ke-Lin Wang$^{4}$}

\address{
$^{1}$ Center for Statistical and Theoretical Condensed Matter
Physics, Zhejiang Normal University, Jinhua 321004, P. R. China  \\
$^{2}$ Department of Physics, Zhejiang University, Hangzhou 310027,
P. R. China \\
$^{3}$Department of Physics, Southwest University of  Science and Technology, Mianyang 621010, P.  R.  China\\
$^{4}$Department of Modern Physics, University of  Science and
Technology of China,  Hefei 230026, P.  R.  China
 }
\date{\today}

\begin{abstract}
Entanglement evolution of two independent Jaynes-Cummings atoms
without rotating-wave approximation (RWA) is studied by an
numerically exact approach. The previous results in the RWA are
essentially modified in the strong coupling regime ($g\ge 0.1$),
which has been reached in the recent experiments on the flux qubit
coupled to the LC resonator. For the initial Bell state with
anti-correlated spins, the entanglement sudden death (ESD) is absent
in the RWA, but does appear in the present numerical calculation
without RWA. Aperiodic entanglement evolution in the strong coupling
regime is observed. The strong atom-cavity coupling facilitates the
ESD. The sign of detuning play a essential role in the entanglement
evolution for strong coupling, which is irrelevant in the RWA. An
analytical results based on an unitary transformation are also
given, which could not modify the RWA picture essentially. It is
suggested that the activation of the photons may be the origin of
the ESD. The present theoretical results could be applied to
artificial atoms realized in recent experiments.
\end{abstract}

\maketitle

\section{introduction}

Quantum entanglement, one of the most striking consequences of nonlocal
quantum correlation, is fundamental in quantum physics both for
understanding the nonlocality of quantum mechanics\cite{Einstein} and its
role in quantum computations and communications\cite{Nielsen}. The
entanglement would undergo decoherence due to the unavoidable interaction
with the environment. As a result, an initially entangled two-qubit system
becomes totally disentangled after evolving for a finite time. This
phenomena is called entanglement sudden death (ESD)\cite{yu} and has been
recently demonstrated experimentally\cite{Almeida}.

Many works are devoted to the ESD of the qubits coupled to an environment
that results in irreversible loss, and the rotating-wave approximation (RWA)
is made on the interaction of the qubits with the field\cite{yu,Bellomo}.
The effect of the counter-rotating terms on the ESD has been less studied,
even for the Jaynes-Cummings (JC) model\cite{JC} where the qubit interacting
only with the single-cavity mode. In fact, some investigations have also
focused on the storing of entanglement in such a system\cite{ye}. So
entanglement dynamics for two independent JC atoms is also of fundamental
interest, and has been well studied only in the RWA\cite{Eberly1,Yonac,Ficek,Sainz}%
.

On the other hand, the JC model is also closely related to condensed matter
physics recently. It can be realized in some solid-state systems recently,
such as one Josephson charge qubit coupling to an electromagnetic resonator
\cite{Wallraff}, the superconducting quantum interference device coupled
with a nanomechanical resonator\cite{Chiorescu,squid}, and the most recently
LC resonator magnetically coupled to a superconducting qubit\cite{exp}. In
conventional quantum optics, the coupling between the two-level "natural"
atom and the single bosonic mode is quite weak, RWA has been usually
employed. With the progress of the fabrication, the artificial atoms may
interact very strongly with on-chip resonant circuits\cite
{Wallraff,squid,Chiorescu,exp}, RWA can not describe well the strong
coupling regime\cite{liu}. Therefore, it is highly desirable to explore the
entanglement dynamics for two independent JC atoms without RWA.

However, due to the consideration of the counter-rotating terms, the
photon number is not conserved, so the photonic Fock space has
infinite dimensions, any solution without RWA is highly nontrivial.
In the recent years, several non-RWA
approaches\cite{chenqh,zheng,liu,Liutao,Amico,Yuyu} has been
proposed in a few contexts. Especially, by using extended bosonic
coherent states, three of the present authors and a collaborator
have solved the Dicke model without RWA exactly in the numerical
sense\cite{chenqh}.

In this paper, we employ a numerically exact technique to solve JC model
without RWA by means of extended bosonic coherent states. The correlations
among bosons are added step by step until further corrections will not
change the results. All eigenfunctions and eigenvalues can be obtained
exactly. Based on the exact solutions to the single JC model, we can study
the entanglement evolution of two JC atoms easily. Analytical results
without RWA based on an unitary transformation are also presented. The paper
is organized as follows. In Sec.II, the numerically exact solution to the JC
model is proposed in detail and analytical results in terms of unitary
transformation are also provided. The numerical results and discussions are
given in Sec.III. The brief summary is presented finally in the last section.

\section{Model Hamiltonian}

The JC model can be written in the basis $\left(
\begin{array}{ll}
\left| \uparrow \right\rangle , & \left| \downarrow \right\rangle
\end{array}
\right) $ where the first index denoting spin-up and the second denoting
spin-down
\begin{equation}
H_{JC}=\frac \Delta 2\sigma _z+\omega a^{+}a+\lambda (a+a^{+})\sigma _x
\label{hamiltonian_JC1}
\end{equation}
where $a^{+}$ and $a$ are the bosonic annihilation and creation operators of
the cavity, $\Delta $ and $\omega $ are the frequencies of the atom and
cavity, $g$ is the atom-cavity coupling constant, and $\sigma _k(k=x,y,z)$
is the Pauli matrix of the two-level atoms. Here we set $\hbar =1$. The
detuning is defined as $\delta =\omega -\Delta $.

In this paper, we describe two approaches to solve the single JC model
without RWA. One is the numerically exact approach in terms of extended
bosonic coherent states, the other is an analytical approach based on a
unitary transformation.

\textsl{Numerically exact approach}.-- In order to facilitate the
calculation, we use a transformed Hamiltonian with a rotation around the $y-$%
axis by an angle $\pi /4$, so $\sigma _z$ $\rightarrow \sigma
_x,\sigma _x\rightarrow -\sigma _z$. This rotation can also be
described formally by the following transformation,
\begin{equation}
H_{JC}^{\prime }=VH_{JC}V^{+}=-\frac \Delta 2\sigma _x+\omega a^{+}a+\lambda
(a+a^{+})\sigma _z,  \label{hamiltonian_JCnew}
\end{equation}
where
\[
V=\frac 1{\sqrt{2}}\left(
\begin{array}{ll}
1 & 1 \\
-1 & 1
\end{array}
\right),
\]
By introducing the new operators
\begin{equation}
A=a+g,B=a-g,g=\lambda /\omega,  \label{newoperators}
\end{equation}
we can observe that the linear term for the bosonic operator is removed, and
only the number operators $A^{+}A $ and $B^{+}B$ are left. Therefore the
wavefunction can be expanded in terms of these new operators as
\begin{equation}
\left| \varphi ^{\prime }(t)\right\rangle =\left(
\begin{array}{l}
\sum_{n=0}^{N_{tr}}c_{1n}\left| n\right\rangle _A \\
\sum_{n=0}^{N_{tr}}c_{2n}\left| n\right\rangle _B\label{wavefunction}
\end{array}
\right).
\end{equation}
For $A$ operator, we have
\begin{equation}
\left| n\right\rangle _A=\frac{A^n}{\sqrt{n!}}\left| 0\right\rangle _A=\frac{%
\left( a+g/\omega \right) ^n}{\sqrt{n!}}\left| 0\right\rangle _A.
\label{Aperators}
\end{equation}
The property of $B$ operator is the same.

The Hamiltonian (2) remains unchanged under the transformation \textit{level}
$1\rightarrow \;$\textit{\ level} $2$ and $\;a^{+}($or $a)\rightarrow
-a{}^{+}($or $-a)$. So we can set $c_{2n}=\pm (-1)^nc_{1n}$ (or $c_{1n}=\pm
(-1)^nc_{2n}$ ). The final more concise wavefunction is supposed as
\begin{equation}
\left| \varphi ^{\prime }(t)\right\rangle =\left(
\begin{array}{l}
\sum_{n=0}^{N_{tr}}c_n\left| n\right\rangle _A \\
\pm \sum_{n=0}^{N_{tr}}(-1)^nc_n\left| n\right\rangle _B \label%
{wavefunction2}
\end{array}
\right).
\end{equation}
The Schr$\stackrel{..}{o}$ dinger equation is
\begin{equation}
\omega (m-g^2)c_m\pm \frac \Delta 2\sum_{n=0}^{N_{tr}}D_{mn}c_n=E^{(\pm
)}c_m,  \label{schrodinger}
\end{equation}
where
\begin{equation}
D_{mn}=\exp (-2g^2)\sum_{k=0}^{\min [m,n]}(-1)^{-k}\frac{\sqrt{m!n!}%
(2g)^{m+n-2k}}{(m-k)!(n-k)!k!}.
\end{equation}
The $l-$th eigenfunction $\left| \varphi ^{\prime (l)}\right\rangle $ is
then obtained, which can also be expressed in the original basis of
Hamiltonian (1) as
\begin{equation}
\left| \varphi ^{(l)}\right\rangle =\left(
\begin{array}{l}
\phi _1^{(l)} \\
\phi _2^{(l)}\label{eigenfunction}
\end{array}
\right) =V^{+}\left| \varphi ^{\prime (l)}\right\rangle.
\end{equation}
The wavefunction at anytime then reads
\begin{equation}
\left| \varphi (t)\right\rangle =\sum_{l=1}^{M_0}h^{(l)}\exp (-iE_lt)\left|
\varphi ^{(l)}(t)\right\rangle ,M_0=2(N_{tr}+1),  \label{eigenfunction_time}
\end{equation}
where $h^{(l)}$ is the coefficient to be determined.

For the two independent JC atoms, the eigenfunction of the total system is
given by
\begin{equation}
\left| \psi \right\rangle =\left| \varphi _1\right\rangle \otimes \left|
\varphi _2\right\rangle,  \label{eigenfunction_total}
\end{equation}
where $\varphi _i$ is the eigenfunction the JC model $i(=1,2)$ with
\begin{equation}
\left| \varphi _i\right\rangle =\left(
\begin{array}{l}
\left| \phi _{i,1}\right\rangle \\
\left| \phi _{i,2}\right\rangle \label{eigenfunction_s}
\end{array}
\right).
\end{equation}
The eigenvalue of the total system is $E=E_1\oplus E_2$. The $j-th$
wavefunction for the total system can then be explicitly expressed
in the basis $\left(
\begin{array}{llll}
\left| \uparrow \uparrow \right\rangle , & \left| \uparrow \downarrow
\right\rangle , & \left| \downarrow \uparrow \right\rangle , & \left|
\downarrow \downarrow \right\rangle
\end{array}
\right) $
\begin{equation}
\left| \psi ^{(j)}\right\rangle =\left(
\begin{array}{l}
\phi _{1,1}^{(l)}\phi _{2,1}^{(k)} \\
\phi _{1,1}^{(l)}\phi _{2,2}^{(k)} \\
\phi _{1,2}^{(l)}\phi _{2,1}^{(k)} \\
\phi _{1,2}^{(l)}\phi _{2,2}^{(k)} \label{wavefunction_j}
\end{array}
\right) ,j=1,2,4(N_{tr}+1).
\end{equation}
The wavefunction of the total system at anytime then reads
\begin{equation}
\left| \psi (t)\right\rangle =\sum_{j=1}^{2M_0}f^{(j)}\exp (-iE_jt)\left|
\psi ^{(j)}\right\rangle ,M_0=2(N_{tr}+1),  \label{wavefunction_t}
\end{equation}
where $f^{(j)}$ is the coefficient to be determined.

The initial two Bell states with anti-correlated and correlated spins, which
are denoted by the Bell state 1 and 2 respectively for convenience, in the
original base are the following if we use the column matrix
\begin{equation}
\left| \psi _{Bell}^{(1)}\right\rangle =\left(
\begin{array}{l}
0\left| 00\right\rangle  \\
\cos \alpha \left| 00\right\rangle  \\
\sin \alpha \left| 00\right\rangle  \\
0\left| 00\right\rangle
\end{array}
\right) ,\;\left| \psi _{Bell}^{(2)}\right\rangle =\left(
\begin{array}{l}
\cos \alpha \left| 00\right\rangle  \\
0\left| 00\right\rangle  \\
0\left| 00\right\rangle  \\
\sin \alpha \left| 00\right\rangle \label{Bell2}
\end{array}
\right) ,
\end{equation}
where $\left| 00\right\rangle $ denotes the photon number state in the two
JC atoms, $f^{(l)}$ is determined by using $\left| \psi (0)\right\rangle
=\left| \psi _{Bell}^{(i)}\right\rangle ,i=1,2$ .

For two-qubit states, entanglement can be quantified by the concurrence\cite
{Wootters}. It can be calculated from the following reduced density matrix
\begin{equation}
\rho =Tr_{ph}(\left| \psi (t)\right\rangle \left\langle \psi (t)\right|
)=\sum_{k,l=1}^{N_0}f^{(k)}f^{(l)}e^{-i(E_k-E_l)t}\Pi,  \label{density}
\end{equation}
where $4\times 4$ matrix $\Pi $ is determined by $\left| \psi
^{(j)}\right\rangle $. The $4$ eigenvalues $\lambda _i$ of the matrix $\rho $
in decreasing order give the entanglement
\begin{equation}
C^{AB}(t)=\max [0,\sqrt{\lambda _1}-\sqrt{\lambda _2}-\sqrt{\lambda _3}-%
\sqrt{\lambda _4}] .  \label{entanglement}
\end{equation}

\textsl{Unitary transformation approach}.-- In order to treat JC model
without RWA analytically, we perform a unitary transformation\cite{zheng} on
Hamiltonian (\ref{hamiltonian_JC1}) to eliminate the counter-rotating wave
term
\begin{eqnarray*}
H^S &=&e^SHe^{-S}, \\
H &=&H_0+H_1,H_0=\frac \Delta 2\sigma _z+\omega a^{+}a,H_1=g(a^{+}+a)(\sigma
_{+}+\sigma _{-}), \\
S &=&\frac{g\xi }\omega (a^{+}-a)(\sigma _{+}+\sigma _{-}),
\end{eqnarray*}
where $\xi $ is a parameter to be determined. The transformed Hamiltonian
can be expanded in terms of $g$ up to the $g^2$ term (higher terms are
neglected), and we have
\[
H^S=H_0+H_1^S+H_2^S+O(g^3),
\]
\begin{equation}
H_1^S=H_1+[S,H_0] \\
=g(\frac \Delta \omega \xi -\xi +1)(a^{+}\sigma _{-}+a\sigma _{+})+g(-\frac
\Delta \omega \xi -\xi +1)(a^{+}\sigma _{+}+a\sigma _{-}),
\end{equation}
\begin{equation}
H_2^S=[S,H_1]+\frac 12[S,[S,H_0]]=-\frac{g^2\Delta }{(\omega +\Delta )^2}%
\sigma _z-\frac{g^2(\omega +2\Delta )}{(\omega +\Delta )^2}.
\end{equation}
If we choose $\xi =\omega /(\Delta +\omega ),$ the counter-rotating wave
term can be eliminated, then we have the following renormalized JC
Hamiltonian in a RWA form
\begin{eqnarray}
H^S &=&\frac{\Delta _{eff}}2\sigma _z+\omega a^{+}a+g_{eff}(a^{+}\sigma
_{-}+a\sigma _{+}),  \label{hamiltonian_ren} \\
\Delta _{eff} &=&\Delta \left( 1-\frac{2g^2}{(\Delta +\omega )^2}\right) ,
\label{d_eff} \\
g_{eff} &=&g\left( \frac{2\Delta }{\omega +\Delta }\right),  \label{g_eff}
\end{eqnarray}
where a constant is removed. The effective detunning can be expressed in
terms of original detunning $\delta $ as
\begin{equation}
\delta _{eff} =\omega -\Delta _{eff}=\left( \delta +\frac{2g^2\left( \omega
-\delta \right) }{\left( 2\omega -\delta \right) ^2}\right).
\label{detunning_eq}
\end{equation}
For $\delta =0,$ we have
\begin{equation}
\delta _{eff}^0=\frac{g^2}{2\omega },\Delta _{eff}^0=\Delta -\delta
_{eff}^0,\;\;g_{eff}^0=g.
\end{equation}
Note that even for zero detunning, we have finite effective detunning after
the unitary transformation. What is more, the non-RWA results within the
unitary transformation approach is expected to be very close to the RWA ones
in the weak coupling regime, because the effective $g$ is not changed, and
only modification is the transition frequency which is reduced by a small
amount proportional to $g^2$.

Follow the derivation for the RWA case in Ref. \cite{Eberly1}, if the
initial state is the Bell state 1, the concurrence for two identical JC
atoms without RWA by the unitary transformation can be easily obtained
\begin{equation}
C_{AB}(t)=\left| \sin 2\alpha \right| \left[ 1-4N^2\sin ^2(\nu t/2)\right],
\label{ana_C_1}
\end{equation}
where
\begin{eqnarray}
N &=&\frac 1{2\sqrt{1+\left[ \frac{\left( \delta +\frac{2g^2\left( \omega
-\delta \right) }{\left( 2\omega -\delta \right) ^2}\right) }{2\left( \frac{%
2g}{1+1/\left( 1-\delta /\omega \right) }\right) }\right] ^2}}, \\
\nu &=&\sqrt{\left( \delta +\frac{2g^2\left( \omega -\delta \right) }{\left(
2\omega -\delta \right) ^2}\right) ^2+4\left( \frac{2g}{1+1/\left( 1-\delta
/\omega \right) }\right) ^2}.
\end{eqnarray}
For zero detunning, we have
\begin{equation}
N_0=\frac 1{2\sqrt{1+\frac{g^2}{16\omega ^2}}},\nu _{_0}=\frac g{N_0}.
\label{ana_N0}
\end{equation}
This is to say, within unitary transformation approach, there is no ESD from
the initial Bell state 1, similar to the RWA case.

Similarly, we can get the entanglement evolution from the initial Bell state
2
\begin{equation}
C_{AB}(t)=\left[ 1-4N^2\sin ^2(\nu t/2)\right] \left( \left| \sin 2\alpha
\right| -8N^2\sin ^2(\nu t/2)\cos ^2\alpha \right),  \label{ana_C_2}
\end{equation}
The ESD occurs for
\begin{equation}
\left| \tan \alpha \right| <4N^2\sin ^2(\nu t/2).
\end{equation}

Note form Eq. (\ref{ana_N0}) that the value of $N_0$ is very close to the
RWA one even for $g=1$, so it is expected that the non-RWA results within
unitary transformation approach is nearly the same as that in RWA in a very
large coupling regime. Moreover, the unitary transformation approach could
not provide essentially different results, because final renormalized RWA
Hamiltonian is derived for all calculations. In our numerically exact
studies, the RWA-type Hamiltonian is not necessary, so some new results may
be obtained.

\section{Results and discussions}

\begin{figure}[tbp]
\includegraphics[scale=0.85]{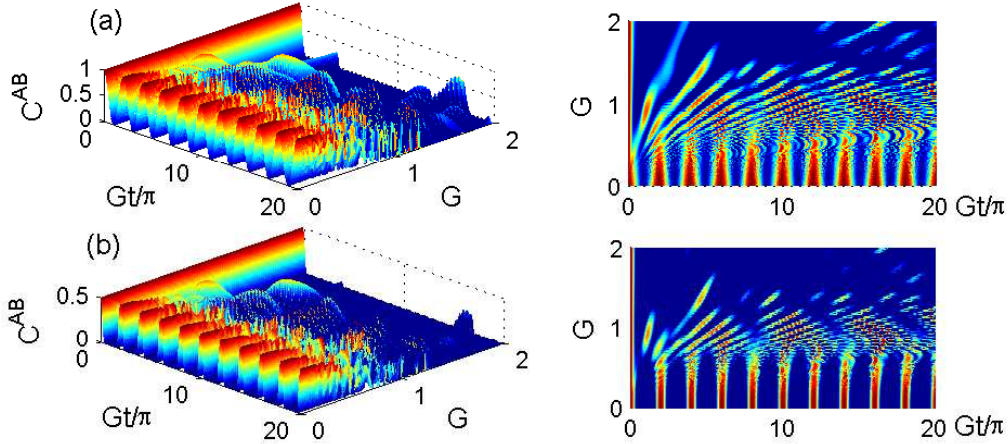}
\caption{ (Color online) The concurrence for atom-atom entanglement with the
initial atomic state: (a) Bell state 1 for $\alpha=\pi/4$ and (b) Bell state
2 for $\alpha=\pi/12$. $G= 2g$. The corresponding right figures are the
bird's-eye view. }
\label{C_ab_g1=g2}
\end{figure}
We first study the evolution of the concurrence for atom-atom
entanglement for two JC atoms without RWA when the states are
initiated with the given Bell states ($\alpha$ is fixed) (
(\ref{Bell2}). Actually, the dynamics of entanglement in bipartite
quantum systems is sensitive to initial conditions\cite{Roszak,yu}

We consider the two same JC atoms for zero detuning. The concurrence for
atom-atom entanglement as a function of time can be calculated with the use
of Eqs. (\ref{wavefunction_j}) and (\ref{wavefunction_t}). As show in Fig.
\ref{C_ab_g1=g2}(a), for small $G=2g$, the concurrence evolve as $%
cos^2(Gt/2) $, just follow that in RWA \cite{Eberly1} for Bell state 1. For $%
G$ above $0.5$, entanglement can drop to zero, which can not happen in the
two JC atoms in RWA for Bell state 1. The entanglement can be revival
irregularly. JC model is just $N=1$ Dicke model. We attribute this aperiodic
entanglement evolution to the quantum chaos in finite Dicke model\cite{Emary}%
. Without RWA, the system is integral, the emergency of the quantum chaos is
impossible, so regular behavior is always observed\cite{Eberly1,Ficek}.

It is interesting that the area of the ESD become larger with
increasing $G$. At very large coupling constant, the revival of
entanglement will not happen, in sharp contrast with those observed
in the RWA. Without RWA, the rule of the transfer of entanglement
between the two-qubit subsystems  derived in
\cite{Yonac,Ficek,Sainz} may not hold due to presence of the
counter-rotating terms. Without RWA, we argue  that there is no
entanglement invariant because the photon number is not conserved.

\begin{figure}[tbp]
\includegraphics[scale=0.6]{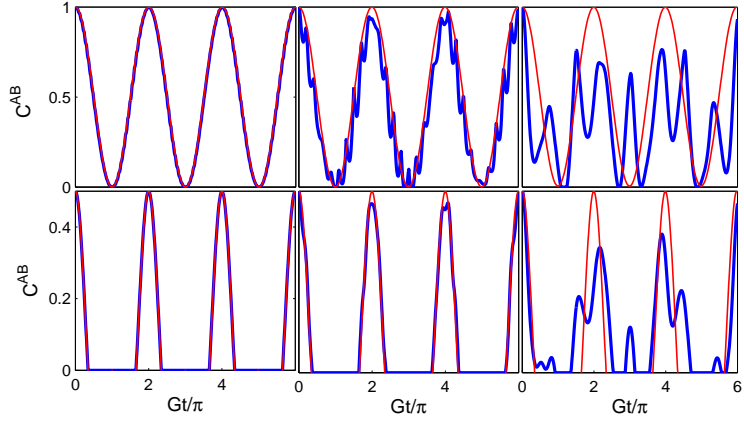} \vspace{-0.8cm}
\caption{(Color online) The concurrence for atom-atom entanglement with the
initial atomic Bell state 1 (upper panel) and the Bell state 2 (down panel).
From left to right column, $g=0.05, 0.1$, and $0.3$. The numerically exact
results and those by the unitary transformation are denoted by the blue and
red curves. }
\label{anaC2tu}
\end{figure}

\begin{figure}[tbp]
\includegraphics[scale=0.85]{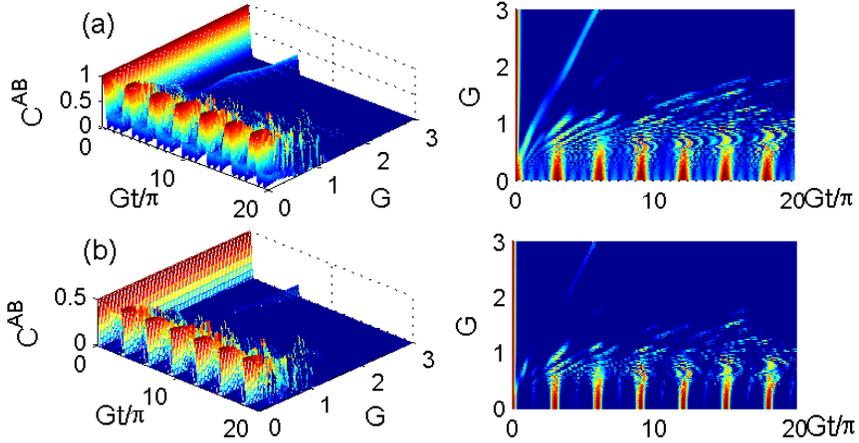}
\caption{ (Color online) The concurrence for atom-atom entanglement with the
initial atomic state: (a) Bell state 1 for $\alpha=\pi/4$ and (b) Bell state
2 for $\alpha=\pi/12$. $g_1=2g_2, G= g_1+g_2$. The corresponding right
figures are the bird's-eye view.}
\label{C_ab_g1g2}
\end{figure}

For Bell state 2, the ESD can happen even in the RWA\cite{Eberly1}.
Without RWA, for small $G$, the entanglement dynamics show the same
behavior. As shown in Fig. \ref{C_ab_g1=g2}(b), the area of the ESD
become larger with increasing $G$, and more wider than that in Bell
state 1.

As we know, for weak coupling between the atom and the single bosonic mode,
the RWA is a good approximation, the non-RWA treatment is not necessary. Our
results for the entanglement dynamics also provide a evidence for this
point. However, for strong coupling case, the entanglement dynamics
demonstrate that the non-RWA should be considered essentially. As the new
progress in the fabrication, some artificial atoms are just strongly coupled
with the bosonic field.

The non-RWA JC model can be also treated by the unitary transformation
approach. Is this approach really good enough in the strong coupling regime?
We now present the analytical results based on this approach for zero
detunning according to Eqs. (\ref{ana_C_1}) and (\ref{ana_C_2}), which are
shown in Fig. \ref{anaC2tu}. The corresponding numerically exact ones are
also exhibited for comparison. It is obvious that the analytical results
essentially deviate from the exact ones at $g= 0.1$. Recent experiments on
the LC resonator coupled to a flux qubit demonstrated that the system
operates in the ultra-strong coupling regime $g=0.1$ \cite{exp}, which
crosses the limit of validity for RWA in the JC model. Since the analysis
based on a unitary transformation could not essentially change the RWA
results, the present numerically exact results are more necessary.

We next consider the two JC atoms with different atom-cavity coupling for
zero detuning. Without loss of generality, we choose $g_1=2g_2$. The
concurrence for atom-atom entanglement as a function of time can also be
calculated with the use of Eqs. (\ref{wavefunction_j}) and (\ref
{wavefunction_t}). As shown in Fig. \ref{C_ab_g1g2}(a), for small $G$, the
concurrence evolve as $cos^2(Gt/2)$, following that in RWA \cite{Eberly1}
for Bell state 1. As $G$ increases, entanglement can drop to zero and is
almost not recovered, quite different from that in the two same JC atoms. We
argue that two JC atoms with different coupling strength suppress the
atom-atom entanglement. This point is confirmed by the evolution of the
initial Bell state 2 as indicated in Fig. \ref{C_ab_g1g2}(b). No matter
whether RWA is taken into account or not, the area of the ESD in the
evolution of the initial Bell state 2 is larger than that in initial Bell
state 1.

\begin{figure}[tbp]
\includegraphics[scale=0.8]{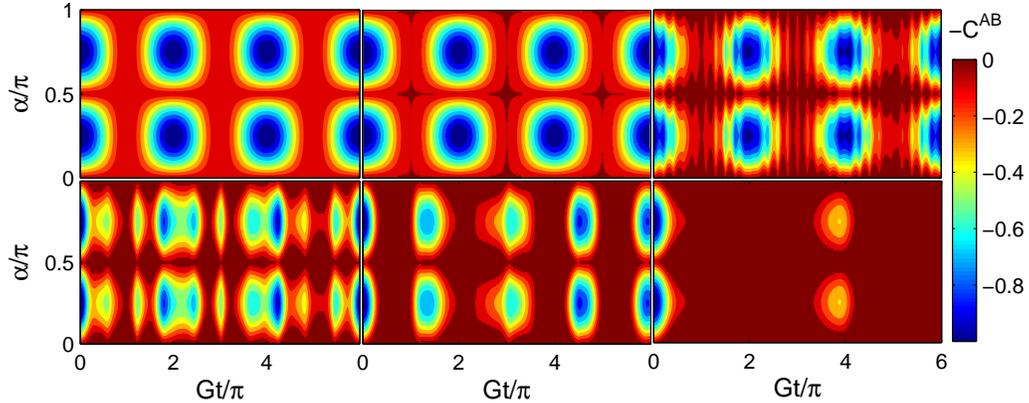}
\caption{ (Color online) Histogram of the concurrence for atom-atom
entanglement with the initial Bell state 1. From left to right, $g=10^{-4},
10^{-2}, 10^{-1}$ (top panel) and $g=0.25, 0.5, 1$ (bottom panel). $G=2g$. }
\label{C_ab_g1}
\end{figure}

\begin{figure}[tbp]
\includegraphics[scale=0.8]{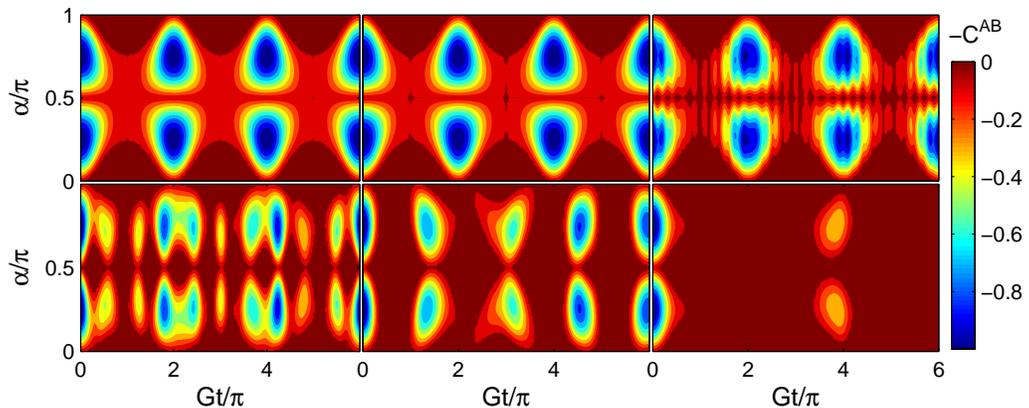}
\caption{ (Color online) Histogram of the concurrence for atom-atom
entanglement with the initial Bell state 2. From left to right, $g=10^{-4},
10^{-2}, 10^{-1}$ (top panel) and $g=0.25, 0.5, 1$ (bottom panel). $G=2g$.}
\label{C_ab_g2}
\end{figure}

For more detail, we study the effect of the coupling strength on evolution
of the concurrence for atom-atom entanglement. Without loss of generality,
we only consider the two same JC atoms. As exhibited in Figs. \ref{C_ab_g1}
and \ref{C_ab_g2}, for weak coupling $g_1<10^{-3}$, no ESD is observed. For $%
g_1>10^{-3}$, the ESD appear, and its area is enlarged with $g$. At the same
time, the periodicity of the entanglement evolution is destroyed.

\begin{figure}[tbp]
\includegraphics[scale=0.8]{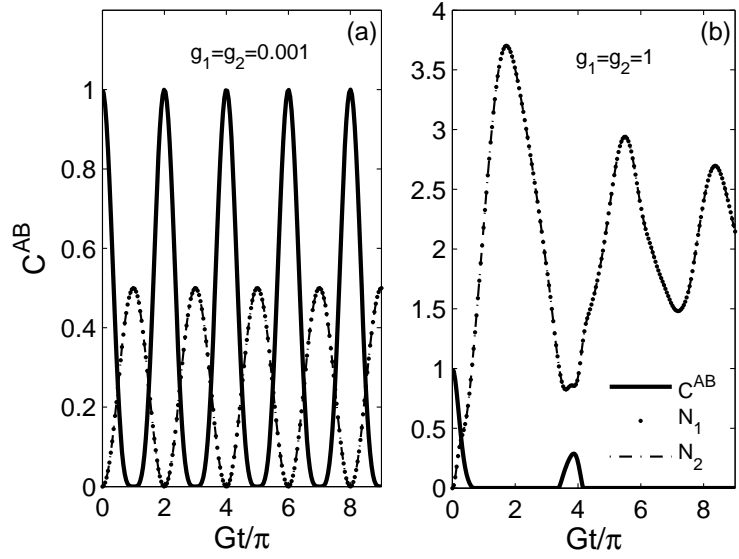}
\caption{ The concurrence for atom-atom entanglement and the averaged
photonic numbers for two identical JC atoms with the initial Bell state 1
for (a) $g=0.01$ and (b) g=1. $G=2g, \alpha=\pi/4$.}
\label{phontonicnumber1}
\end{figure}

\begin{figure}[tbp]
\includegraphics[scale=0.8]{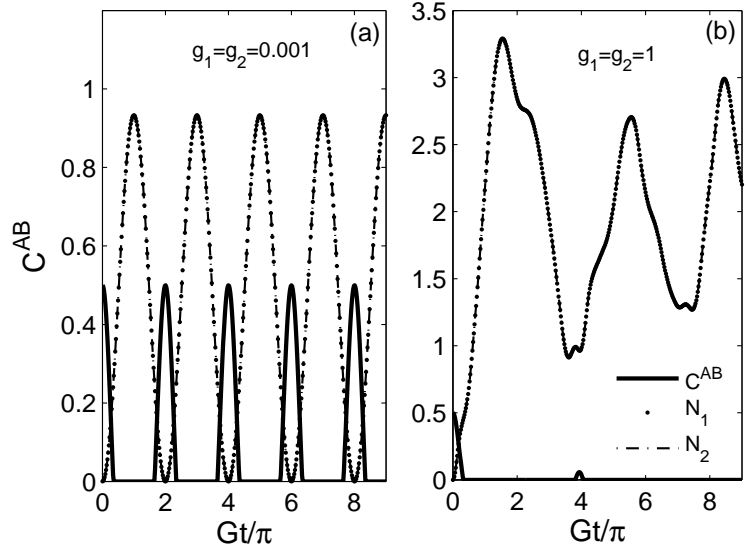}
\caption{ The concurrence for atom-atom entanglement and the averaged
photonic numbers for two identical JC atoms with the initial Bell state 2
for (a) $g=0.01$ and (b) g=1. $G=2g, \alpha=\pi/12$.}
\label{phontonicnumber2}
\end{figure}

One natural question is what the mechanism is for the ESD in two
remote JC models. The photons should influence the atomic state
considerably. We then calculate the average photon number during the
evolution. To show its role in  the entanglement dynamics, we plot
the average photon number together with entanglement for two limit
coupling cases $g=0.001$ and $1$ in Fig. \ref{phontonicnumber1} with
initial Bell state 1 and Fig. \ref {phontonicnumber2} with initial
Bell state 2. It is very interesting that the curves for the
entanglement and the average photon number always show opposite
behavior. It is highly suggested that the average photon number
suppress the entanglement between two atoms considerably and
sensitively for both weak and strong coupling.

\begin{figure}[tbp]
\includegraphics[scale=0.85]{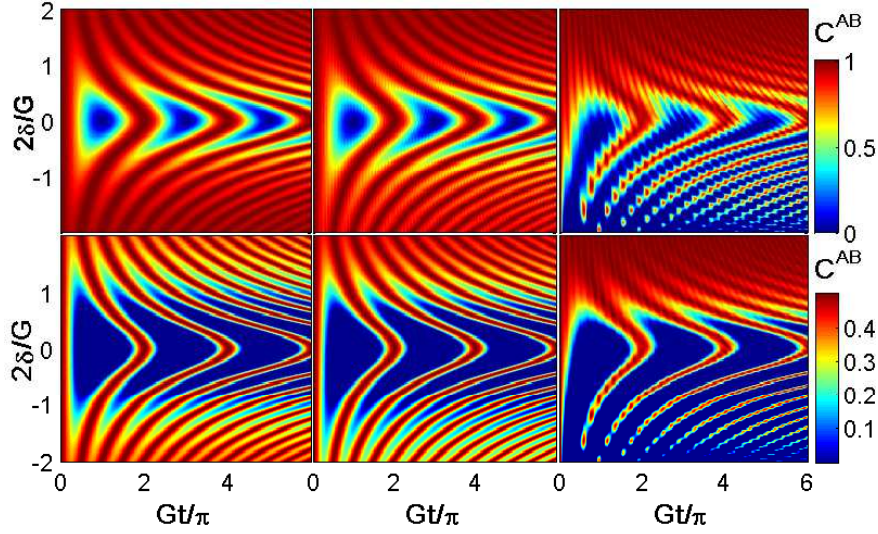}
\caption{ Effect of detuning on the entanglement evolution of the Bell
initial state 1 with $\alpha=\pi/4$(top panel) and Bell initial state 2 with
$\alpha=\pi/12$(bottom panel). From left to right column, $g=10^{-4}, 0.02,
0.1$. $\delta =\omega -\Delta, G=2g$.}
\label{detunning}
\end{figure}

Finally, we turn to the effect of detuning on the entanglement
evolution. It is found in Ref. \cite{Ficek} that the atom-atom
entanglement in RWA usually increases with ratio between detuning
and coupling constant $\left| \delta \right| /g$ and is irrelevant
with the sign of detuning. In other words, the entanglement
decreases with $g$ for given detuning. Without RWA, it is not that
case. After the unitary transformation, we have the renormalized JC
Hamiltonian Eq. (\ref{hamiltonian_ren}) in a RWA type. So
entanglement evolution depends on $\left| \delta _{eff}\right| $,
which is related to both the magnitude and the sign of original
detuning $\left| \delta \right| $ shown in Eq. (\ref{d_eff}). By
Eqs. (\ref{g_eff}) and (\ref{detunning_eq}), the effective coupling
constant can also be expressed as $g_{eff}=2g/\left[ 1+1/\left(
1-\delta /\omega \right) \right] $ if the counter rotating-wave
terms are taken into account after the unitary transformation. It
follows that $g_{eff}$ decreases with $\left| \delta \right| $ for
positive detuning and increases with $\left| \delta \right| $ for
negative detuning. So in the strong coupling regime where the
counter rotating-wave terms is required, the effect of detuning is
very sensitive to the sign of detuning. The above discussions should
be also suited qualitatively to the numerical exact results. We
calculate the entanglement evolution of the initial Bell  states 1
and 2 as a function of detuning. The results for different coupling
constants ranging from weak to strong coupling regime are list in
Fig. \ref {detunning}. The entanglement evolution is symmetric with
detuning for the weak coupling and becomes asymmetric with the
increase of the coupling constant. In the weak coupling regime, as
shown in the left column, the entanglement increase with the value
of detuning, and independent of the sign, similar to those in RWA.
With the increase of the coupling, it is however observed that the
entanglement decreases with the magnitude of the positive detuning,
and increases with the magnitude of the minus detuning, in contrast
to the case of RWA. In the strong coupling regime, the positive
detuning stabilizes the entanglement and facilitates its revival,
whereas the negative detuning suppresses the entanglement,
facilitates its death, and reduces the period of entanglement
revival.

\section{Conclusions}

In this paper, based on the exact solution for the single JC atoms, we are
able to calculate exactly the entanglement evolution of the two independent
JC atoms without RWA. The results are essentially different from RWA cases
in the strong coupling regime. The analytical results based on an unitary
transformation are also given. It can modify the RWA results and could not
provide an essentially different ones. Initiated from the Bell state 1, the
RWA results show no ESD. The present numerically exact calculations for the
non-RWA model show that the ESD could not be avoided, and the periodicity of
entanglement evolution is destroyed by the presence of additional photons.
We also suggest that the photons may suppress the entanglement and is just
the origin of the ESD. The effect of the detunning on the entanglement
evolution is also investigated. It is observed that the sign of detunning
play a essential role in the strong coupling regime. The present theoretical
prediction would be tested in a experimental study of ESD where the
artificial atoms are made of circuit QED\cite{Wallraff,squid,Chiorescu,exp}
if operating in the ultra-strong coupling regime.

\section*{ACKNOWLEDGEMENTS}

This work was supported by National Natural Science Foundation of
China and National Basic Research Program of China (Grant Nos.
2011CB605903 and 2009CB929104).

\end{document}